%% file: main.tex
\documentclass{IEEEtran}
\usepackage{cite}
\usepackage{amsmath,amssymb,amsfonts,upgreek}
\usepackage{graphicx}
\usepackage{subcaption}
\usepackage{textcomp,nicefrac}
\usepackage{array}
\usepackage[mathlines,switch]{lineno}
\usepackage{eso-pic}

\def\BibTeX{{\rm B\kern-.05em{\sc i\kern-.025em b}\kern-.08em
T\kern-.1667em\lower.7ex\hbox{E}\kern-.125emX}}
\AddToShipoutPictureFG*{%
  \AtPageLowerLeft{%
    \hspace*{\dimexpr1in+\oddsidemargin\relax}%
    \raisebox{0.5cm}{%
      \parbox{\textwidth}{\footnotesize\centering
      © 2026 IEEE. Personal use of this material is permitted. Permission from IEEE must be obtained for all other uses, in any current or future media, including reprinting/republishing this material for advertising or promotional purposes, creating new collective works, for resale or redistribution to servers or lists, or reuse of any copyrighted component of this work in other works.
      }
    }
  }
}

\begin{document}
\bstctlcite{IEEEexample:BSTcontrol}
\title{A Novel Approach for Fault Detection and Failure Analysis of CMOS Copper Metal Stacks}

\include{ieee_author_block.tex}

\maketitle

\begin{abstract}
 
For the Inner Tracking System 3 (ITS3) upgrade, the ALICE experiment at CERN requires monolithic active pixel sensors of dimensions up to 97~mm$\,\times\,$266~mm, occupying a large fraction of a 300 mm wafer. To manufacture such a wafer-scale device, larger than the single design reticle size, stitching is employed. The MOnolithic Stitched Sensor (MOSS) is a prototype silicon pixel sensor of 14~mm$\,\times\,$259~mm size with the primary goal of understanding the stitching technique and yield. Given the large size, high yield is paramount for the ITS3 sensors, and an in-depth yield characterization was performed on these MOSS sensors. In a collaborative effort, the foundry adapted the metal stack to the requirements of the project, but recurrent fault signatures were discovered with various frequencies across all 20 wafers tested, and correlated through dedicated measurements and analyses. Following these findings, the foundry implemented a mitigation strategy to avoid the issue in the future. This article does not describe process details but concentrates on the measurements and analysis method.

\end{abstract}


\section{Introduction}
\label{sec:introduction}
\IEEEPARstart{T}{he} Inner Tracking System 3 will replace the innermost three tracking layers of the ALICE experiment at the LHC at CERN~\cite{ALICE_2008, Upgrade_ITS_2013, TDR_2024}. The ITS3 is based on cylindrically bent, wafer-scale, monolithic active pixel sensors manufactured in 65~nm CMOS technology. Stitching is employed to manufacture sensor layers of up to 97~mm$\,\times\,$266~mm size on 300~mm (12~in) wafers, far exceeding the maximum single design reticle dimensions of about 25~mm$\,\times\,$32~mm~\cite{Stitching_1,tower-stitching}. High yield is therefore crucial for the successful fabrication of sensor layers. A prototype MOnolithic Stitched Sensor (MOSS) of 14~mm$\,\times\,$259~mm size was developed to study the feasibility of the stitching process and yield requirements for this application. A set of 24 wafers with 6 MOSS sensors each was produced, and tests were performed on chips from up to 20 wafers.

To extract maximum information from the limited chip sample, a step-wise power-up procedure was introduced. Prior to powering the chip, impedance measurements are performed. The power nets are then ramped up to nominal voltage, while currents are recorded, and the chip is monitored with a thermal camera. Faults are identified as low impedances, ohmic turn-on currents, and coincident thermal hotspots. Correlating this information with the chip layout allows for the identification of potential process or design issues. Statistical analysis was used to further pinpoint the issue, combined with Focused Ion Beam (FIB) cross-sectioning.
The presented method allows for early fault detection and fast feedback in collaboration with the foundry and chip designers.  

In Section II, the MOSS chip is described. Section III presents the measurement techniques used in this study. Section IV discusses the data analysis results. Statistical and FIB validation are discussed in Section V, along with mitigation and testing strategy, before conclusions are drawn in Section VI.

\section{MOSS prototype sensor}

\begin{figure}[t]
\centerline{\includegraphics[width=3.55in]{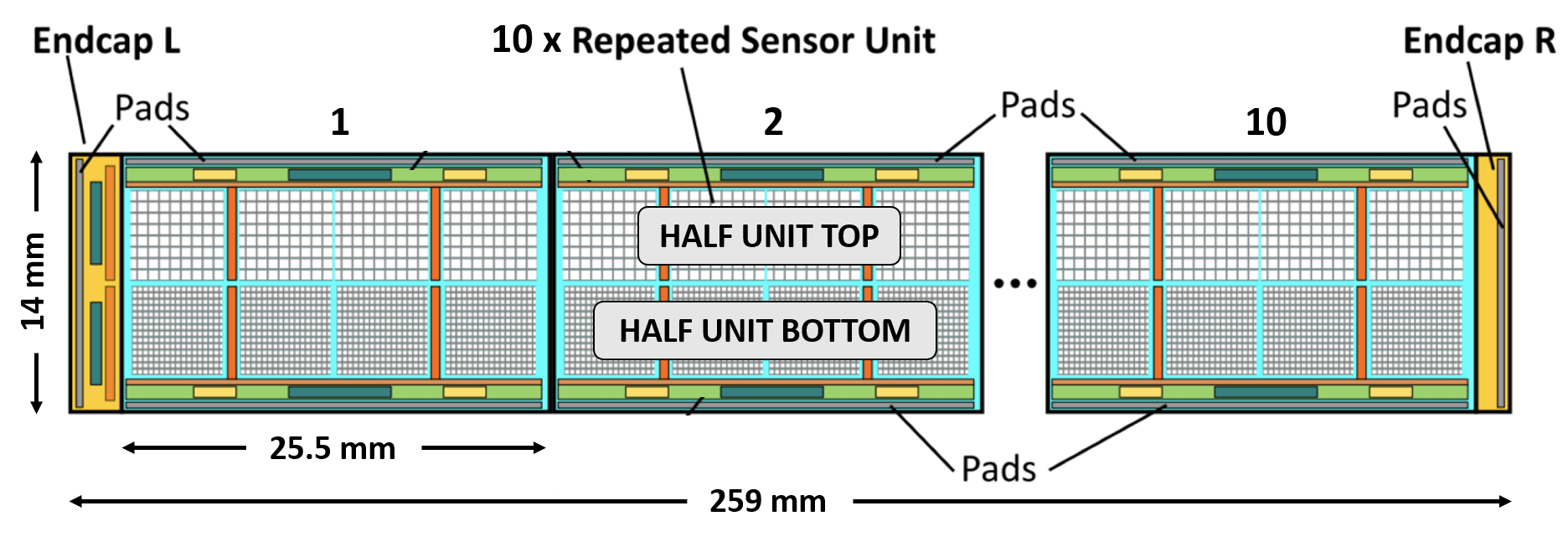}}
\caption{Schematic structure of the MOSS sensor.}
\label{MOSS_stitch}
\end{figure}

\begin{figure}[t]
\centerline{\includegraphics[width=3.35in]{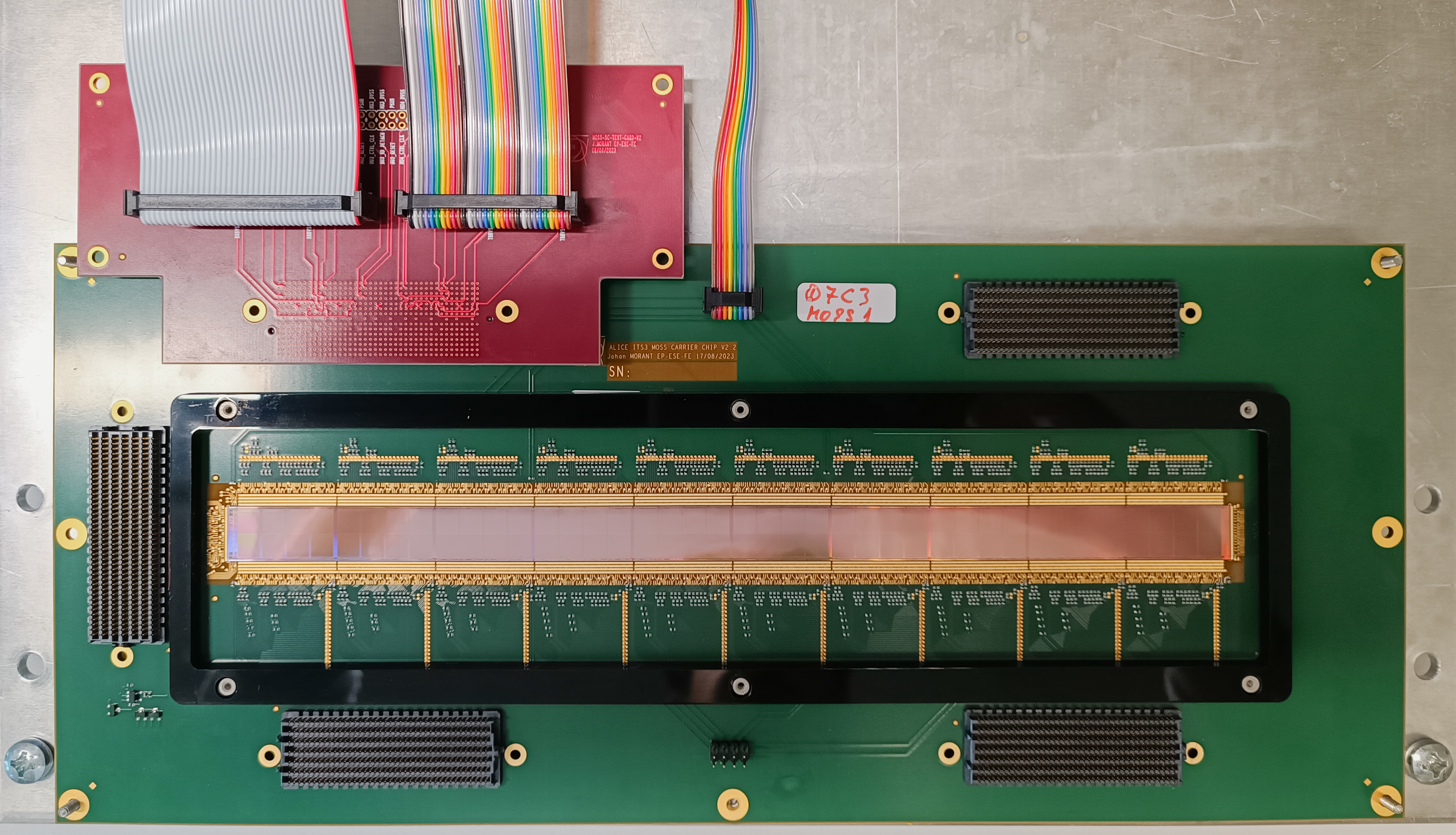}}
\caption{MOSS sensor (center) mounted on the testing PCB (green), with the protective cover removed. A breakout board (red) in impedance measurement configuration is connected to the top left high-density connector. The dimensions of the testing PCB are 150~mm$\,\times\,$350~mm.}
\label{MOSS_imp_breakout}
\end{figure}

The MOSS chip is shown in Fig.~\ref{MOSS_stitch}. Ten identical repeated sensor units (RSU) are stitched together on the same die, creating one monolithic sensor with the left and right endcaps completing the design structure. Each RSU comprises 4 pixel matrices in the top Half Unit (HU), and 4 pixel matrices in the bottom HU, with each 256$\,\times\,$256 pixels and 320$\,\times\,$320 pixels per matrix, respectively~\cite{MOSS_2023}. For the future ITS3 sensors, the chips will be read out and powered exclusively from the left and right endcaps. In the MOSS design, each of the 20 HUs is interfaced via an individual set of wire bond pads along the long edges of the chip, featuring independent power domains. This design granularity allows the characterization and operation of sub-structures of the MOSS sensor in case of faults on other parts of the chip. There are 8 power nets for each HU. Using custom-developed tools and procedures, the sensor is mounted on a purely passive testing PCB. In total, 2192 wire bonds electrically interconnect the chip to the PCB. Four high-density connectors, each with 560 pins, allow for connecting to 5 HUs each. A fifth connector on the left edge of the test PCB is wired to the left endcap structure of the MOSS sensor, used for testing the stitched communication bus. It is not used for powering and testing individual HUs. A breakout board (see Fig.~\ref{MOSS_imp_breakout}) gives access to the 8 power nets for each HU. Table~\ref{tab:power_nets} describes the naming and associated functional domain of the power nets. The Backbone domain spans the full length of the MOSS sensor, crossing the stitching boundaries between each RSU. 

The MOSS sensor is manufactured in 65~nm CMOS technology, with a dual-damascene copper metal stack \cite{GAMBINO2012221}.

\begin{table}[h]
    \centering
    \caption{\textsc{MOSS Power Domains}}
    \begin{tabular}{| >{\centering\arraybackslash}m{1.3cm} | >{\centering\arraybackslash}m{1.3cm} | >{\centering\arraybackslash}m{2.5cm} | >{\centering\arraybackslash}m{1.5cm} |}
         \hline
         Ground net & Supply net & Functional domain & Nominal voltage [V]\\
         \hline 
         \hline
         AVSS & AVDD & Analog & 1.2\\
         DVSS & DVDD & Digital & 1.2\\
         DVSS & IOVDD & Digital input/output & 1.8\\
         BBVSS & BBVDD & Backbone & 1.2 \\
         PSUB & & Chip substrate & 0 or $-$1.2 \\
         \hline

    \end{tabular}
    \label{tab:power_nets}
\end{table}

\section{Measurement techniques}
Two custom test setups are used to perform the impedance and powering measurements, each described below.
\subsection{Impedance measurement}
\begin{figure}[htb]
    \centering
    \begin{subfigure}[b]{0.49\columnwidth}
    \centering
    \includegraphics[width=0.995\columnwidth, keepaspectratio]{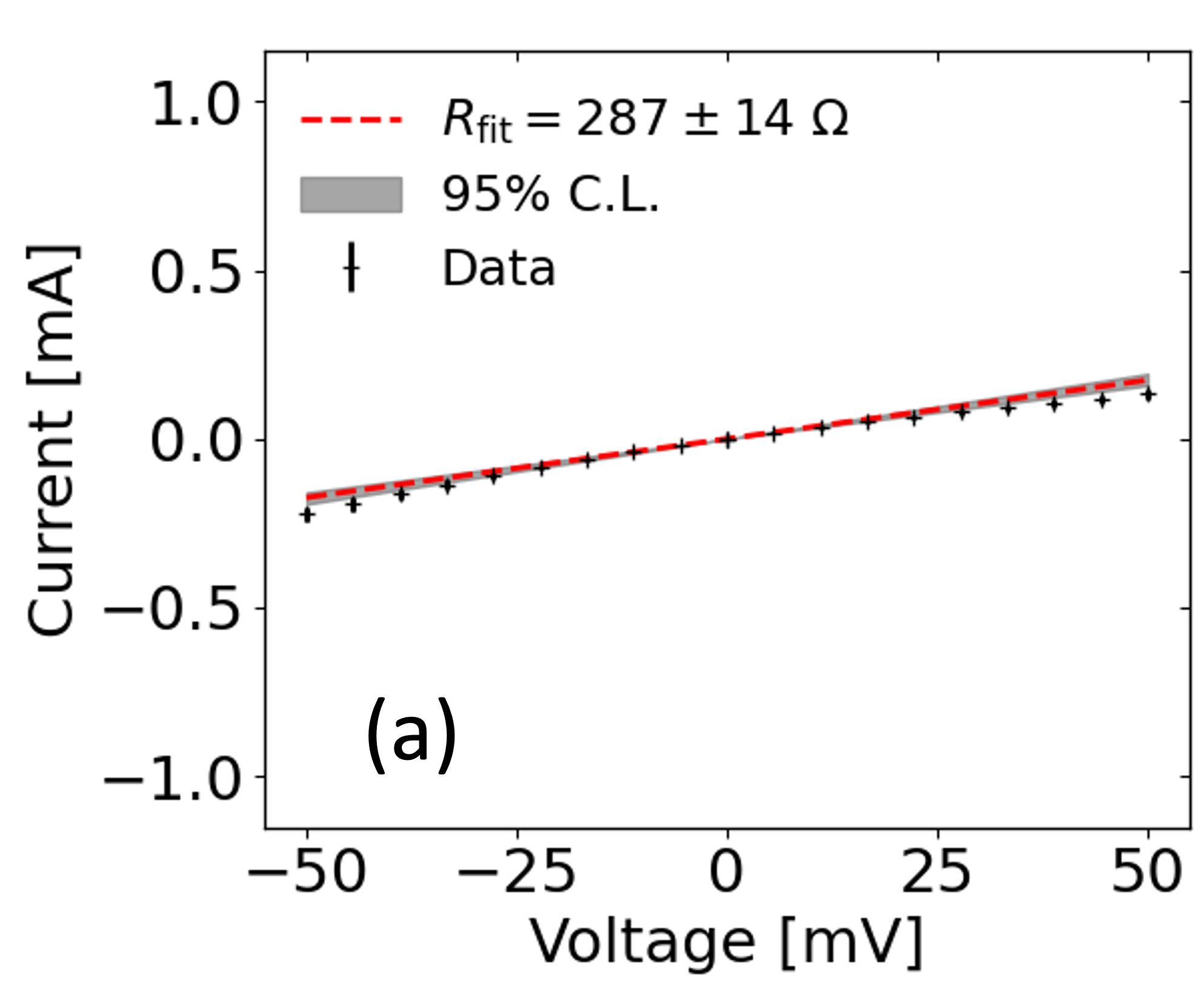}
    \label{fig:Impedance_OK}
    \end{subfigure}
    \hspace{0mm}%
    \begin{subfigure}[b]{0.49\columnwidth}
    \centering
    \includegraphics[width=0.995\columnwidth, keepaspectratio]{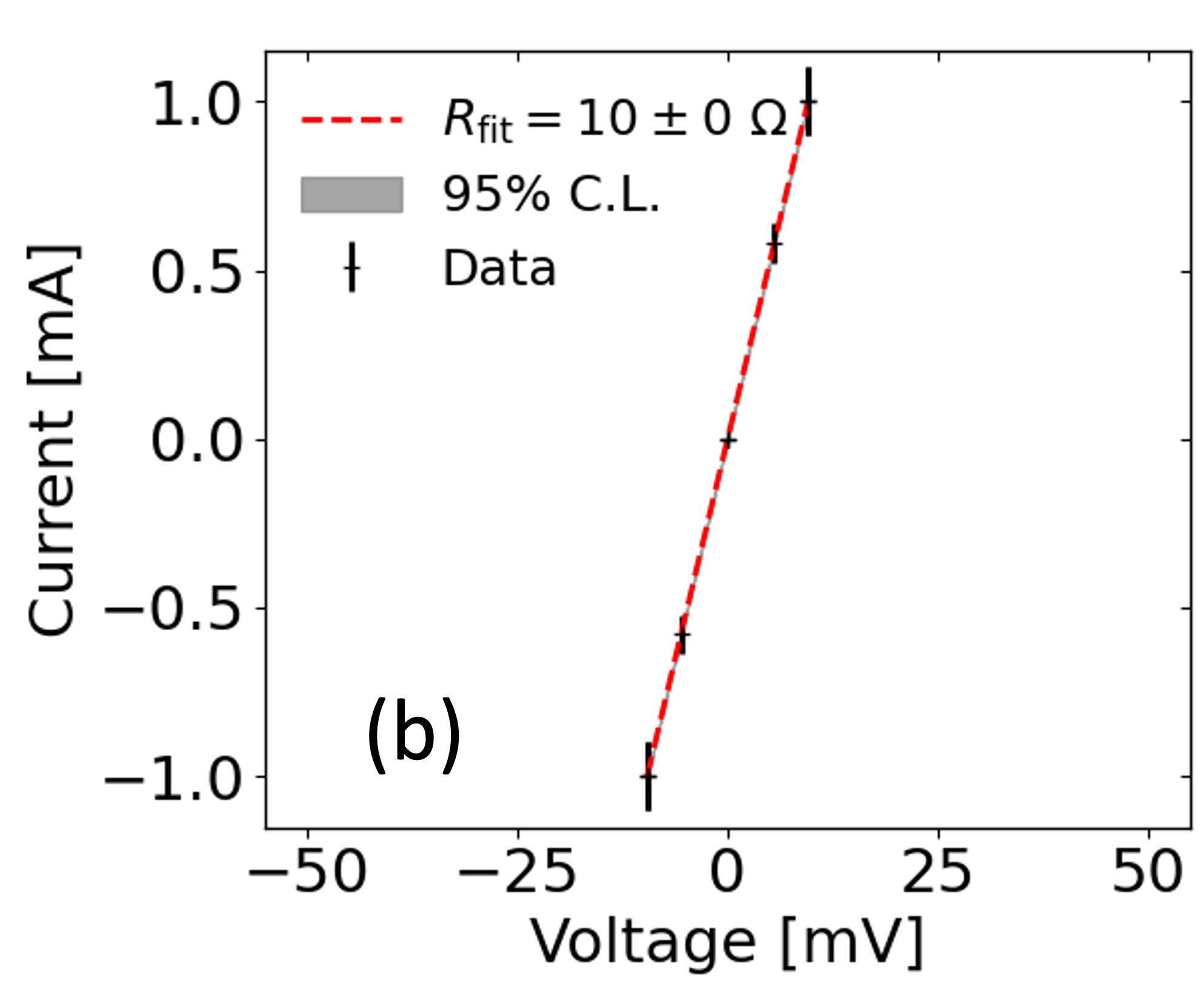}
    \label{fig:Impedance_nOK}
    \end{subfigure}
    \vspace{-1cm}
    \caption{Power net pair succeeding (a), and classified as short (b) in the impedance measurement. }
    \label{fig:Impedance_all}
\end{figure}

Measuring the impedance across all power net pair combinations allows classification of each pair as either `ok' or `short'. The impedance measurement setup consists of two main components: a channel multiplexer and a source measurement unit. The channel multiplexer enables the automatic setting of all $\binom{8}{2}=28$ power net pair combinations for each of the 5 HUs connected via a single breakout board. For every combination, a voltage is applied in steps of 5~mV from 0~to~$-50$~mV and 0~to~$+50$~mV. In this voltage range, transistor and diode structures do not become highly conductive. The current is measured, and the voltage ramp is stopped if the current exceeds 1~mA. A linear fit is performed to approximate the resistance between the nets under test from Ohm's law $U=R\cdot I$. Errors are estimated from the fluctuating laboratory temperature, and lead wire resistances. From the distribution of all measured resistances, an empirical global cut is made at 30~$\Omega$ at the minimum
between the first tail and the following rise in the distribution. For a lower resistance value, the power net pair is deemed to have a short. Fig.~\ref{fig:Impedance_all}a and Fig.~\ref{fig:Impedance_all}b show two examples of net pair measurements classified as `ok' and `short', respectively.

\subsection{Powering setup with thermal camera}

\begin{figure}[b]
\centerline{\includegraphics[width=3.5in]{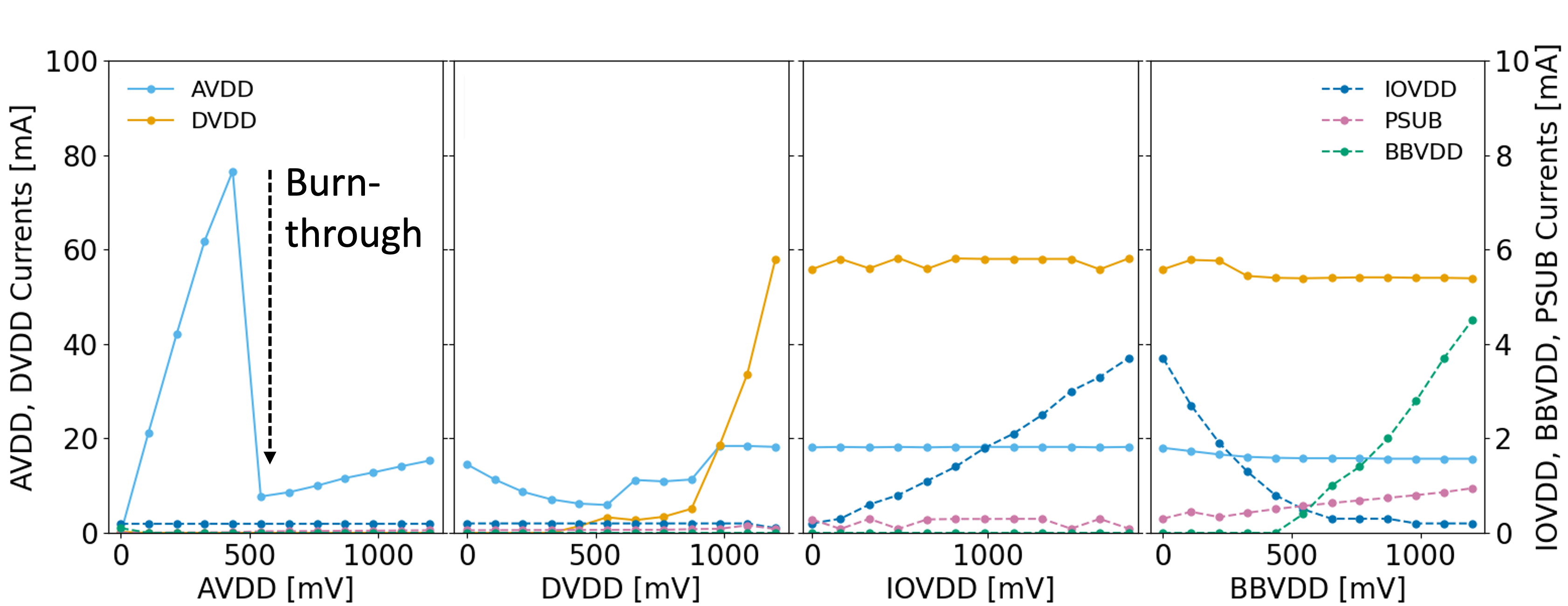}}
\caption{Example of power ramp turn-on curves. The plot on the left shows a burn-through during the ramp-up of the AVDD power net. After the burn-through, the turn-on curves follow the standard behavior of a MOSS chip. The legend entries correspond to the power domains in Table~\ref{tab:power_nets}.}
\label{fig:power_ramp}
\end{figure} 

In the power ramping stage, the turn-on current curves provide information on ohmic turn-on behavior and burn-through events, and a thermal camera is used to locate faults via heat signatures. Each power net is brought to nominal voltage individually for each HU. The chip-specific power-up sequence is shown for one HU in Fig.~\ref{fig:power_ramp}. Here, the first power net (AVDD) is ramped up in steps of 100~mV, while the remaining power nets are held at 0~V. Currents on all power nets are measured. In this example, a steep, ohmic turn-on, corresponding to a short, is observed during ramp-up of the AVDD net. The sharp drop in current indicates a so-called `burn-through' of the short -- visible with a thermal camera as a disappearing hotspot as discussed below -- and the turn-on curve again follows the expected shape. After reaching nominal voltage, the AVDD net is kept powered on, and the DVDD power net is ramped up. This pattern is repeated until all power nets are at nominal voltage. If the currents on any of the power nets exceed an individual current limit, the power ramp is stopped. Because the powering setup does not allow configuration of the chip by writing to multiple registers, the currents in the powered-on state vary. The typical ranges are given in Table~\ref{tab:power_on_currents} for reference at PSUB~=~0~V.

\begin{table}[h]
    \centering
    \caption{\textsc{MOSS Endpoint Power-On Currents}}
    \begin{tabular}{| >{\centering\arraybackslash}m{1.5cm} | >{\centering\arraybackslash}m{1.5cm} | >{\centering\arraybackslash}m{1.7cm} | >{\centering\arraybackslash}m{1.7cm} |}
         \hline
         DVDD [mA] & AVDD [mA] & IOVDD [mA] & BBVDD [mA]\\
         \hline 
         \hline
         1.5 -- 80.0 & 2.0 -- 25.0 & 0.2 -- 0.4 & 2.0 -- 20.0\\
         \hline
    \end{tabular}
    \label{tab:power_on_currents}
\end{table}

\begin{figure}[h]
\centerline{\includegraphics[width=\linewidth]{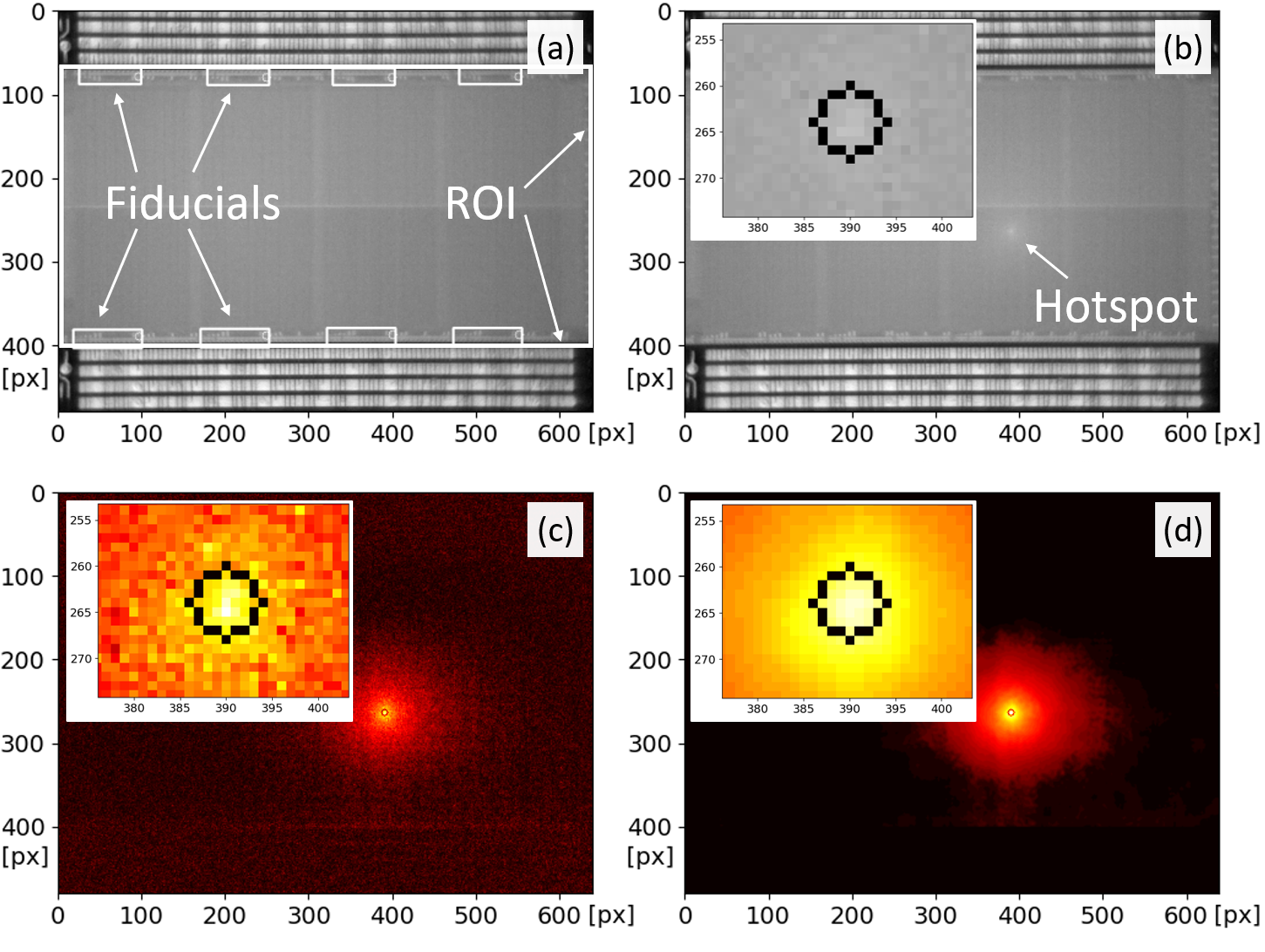}}
\caption{Thermal camera image analysis steps. (a) The HU under test prior to powering with marked fiducials and ROI. (b) The hotspot candidate image. (c) The difference between (a) and (b). (d) The final denoised image used to extract the hotspot location. Insets in (b), (c), (d) illustrate the extracted location. The black circle has a radius of 5 pixels, centered on the hotspot location.}
\label{fig:hotspot}
\end{figure}

The power ramp is performed in a light-shielded box, and a thermal camera with 640$\times$480 pixels and mounted on a motorized linear stage is placed over the HU under test. The resulting resolution of the thermal camera image is 50~$\upmu$m. The field of view covers one RSU. Hotspots correlating to shorts are visualized as shown in Fig.~\ref{fig:hotspot}, corresponding to the peak current during a burn-through (e.g. as in the AVDD power ramp in Fig.~\ref{fig:power_ramp}). The hotspot locations are extracted in a semi-automated fashion:
\begin{enumerate}
    \item One image $I_0$ is taken before powering up the chip. Eight fiducial structures on the chip are located (see Fig.~\ref{fig:hotspot}a), and an affine transformation matrix is estimated by a least squares fit of the fiducial positions and the target positions of the global MOSS sensor design coordinate system. A Region Of Interest (ROI) cut is made on the active chip area, excluding the PCB structures.
    \item Hotspots are identified by following algorithm: For each of the $n$ images $I$ acquired during the power ramp (for a completed ramp $n\simeq520$ at a typical imaging frequency of 6.67~Hz), a difference image is computed between the initial (non-powered) image $I_0$ and image $I$. The ROI cut, an empirical threshold (setting pixels below to 0), and a median blur operation (reducing salt-and-pepper noise) are applied to each difference image, creating $n$ new images ${I}_{ROI}$. From each new image \( {I}_{ROI} \), the average  \( \overline{I}_{ROI} \) is calculated, and subtracted from the average value of a 5$\times$5 pixel mask \( \overline {M}_{5\times5}(x, y, {I}_{ROI}) \) which is scanned over the same new image. Looping over all $n$ images the same way, the image which maximizes the difference between the full ROI average and the 5$\times$5 pixel mask average is taken as a candidate \( I_{\text{cand}} \) (see Fig.~\ref{fig:hotspot}b) for further hotspot analysis: 

    \[ I_{\text{cand}} = \underset{I}{\text{argmax}} \left( \max_{x, y \in ROI} \left|  \overline{I}_{ROI} -  \overline {M}_{5\times5}(x, y, {I}_{ROI}) \right| \right) \]
    \item The simple difference image between the initial image (Fig.~\ref{fig:hotspot}a) and the selected hotspot candidate \( I_{\text{cand}} \) (Fig.~\ref{fig:hotspot}b) is shown in Fig.~\ref{fig:hotspot}c. After a non-local means denoising step is applied to the corresponding image $I_{ROI}$ (improving localisation accuracy, see Fig.~\ref{fig:hotspot}d), the hotspot locations are extracted as the enclosed contour maximum (or center of gravity), and manually confirmed. The extracted hotspot location is indicated as a black circle in the insets of Fig.~\ref{fig:hotspot}. The transformation determined in step 1) is applied to the selection, and the hotspot coordinates are stored. 
\end{enumerate}
Cases of multiple simultaneous hotspots exist, where each hotspot is treated independently. The best achievable resolution window is 50~$\upmu$m$\,\times\,$50~$\upmu$m. On average, a hotspot localization accuracy of $\lesssim\,$100~$\upmu$m can be expected, accounting for a slight barrel distortion at the edges of the image of 1 pixel, and cases of hotspots saturating more than 1 pixel.

\subsection{Chip design correlation}

\begin{figure}[tb]
\centerline{\includegraphics[width=3.in, trim={0cm 0 0 0.2cm},clip,keepaspectratio]{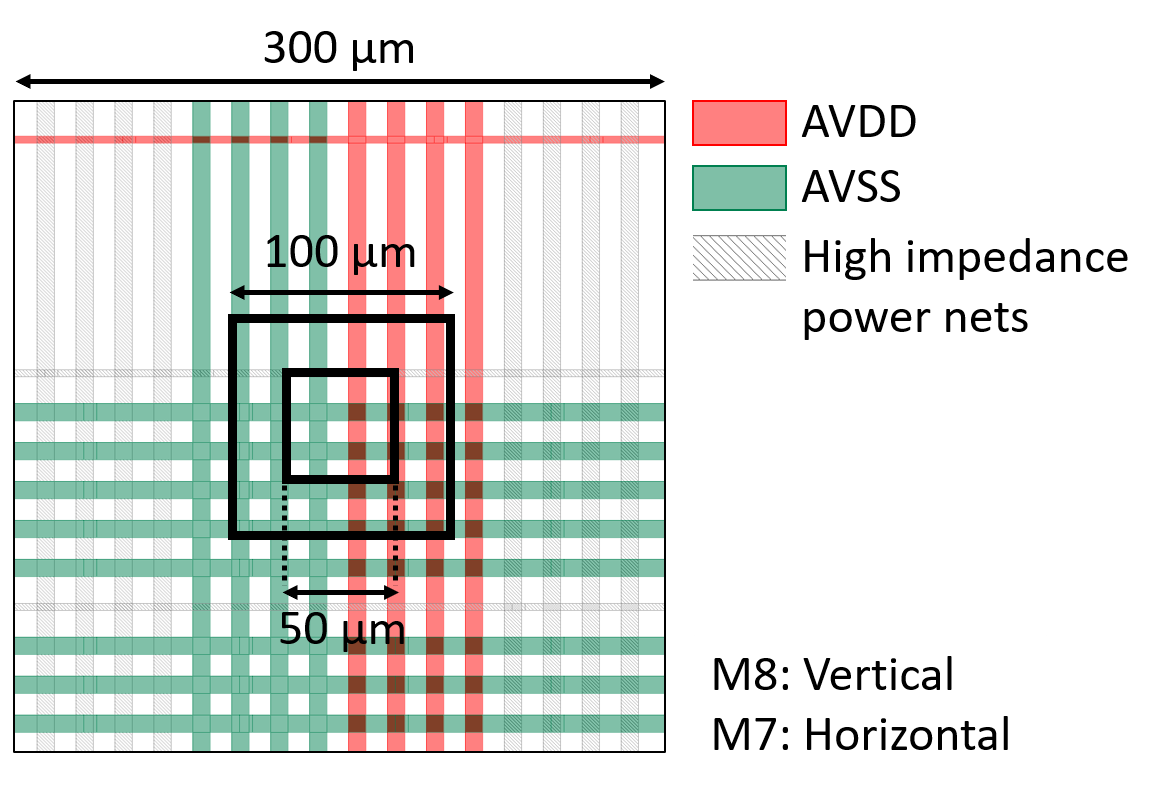}}
\caption{Detail of the MOSS power grid, overlaid with the best case (50~$\upmu$m$\,\times\,$50~$\upmu$m) and average case (100~$\upmu$m$\,\times\,$100~$\upmu$m) resolution windows of the thermal camera. The AVDD and AVSS nets, exhibiting a short, are highlighted in color.}
\label{fig:correl}
\end{figure}

To determine failure-relevant design features, hotspot locations are correlated with the metal stack from the chip design and power nets from the impedance measurement. For a single hotspot and single net pair classified as short, an unambiguous correlation is made. An example overlay of the best case (50~$\upmu$m$\,\times\,$50~$\upmu$m) and average (100~$\upmu$m$\,\times\,$100~$\upmu$m) resolution windows, with the two power nets exhibiting a short highlighted in red and green, is shown in Fig.~\ref{fig:correl}. Only the uppermost copper layers of the chip metal stack (M7, M8) are shown in the Figure. An M7 and M8 metal presence, as well as the presence of affected power nets within the best and average case resolution windows, is extracted from the analysis.

\section{Data analysis}
\subsection{Impedance measurement}
\begin{figure}[t]
\centerline{\includegraphics[width=2.75in]{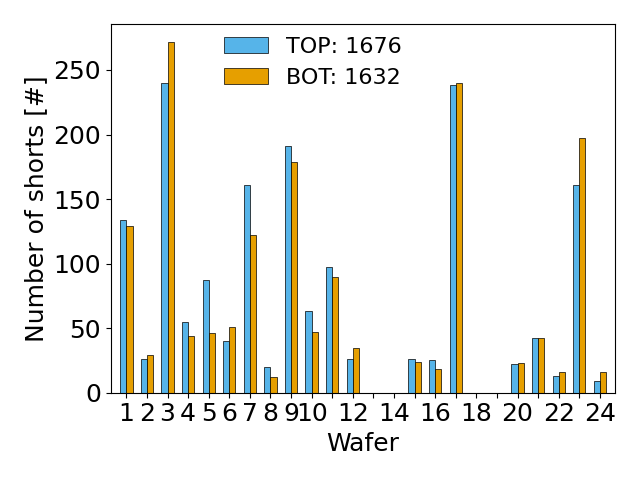}}
\caption{Shorts per wafer, split into the top and bottom half of the MOSS sensors. A strong wafer-to-wafer fluctuation is observed. Wafers 13, 14, 18, 19 were physically damaged and could not be tested.}
\label{fig:short_per_wafer}
\end{figure}

A summary plot showing the number of shorts per wafer, split into top and bottom halves of the MOSS chips, is given in Fig~\ref{fig:short_per_wafer}. Strong fluctuations between wafers are observed, with an even split in the number of shorts between the top and bottom halves of the chip.

\begin{figure}[h]
\centerline{\includegraphics[width=2.2in]{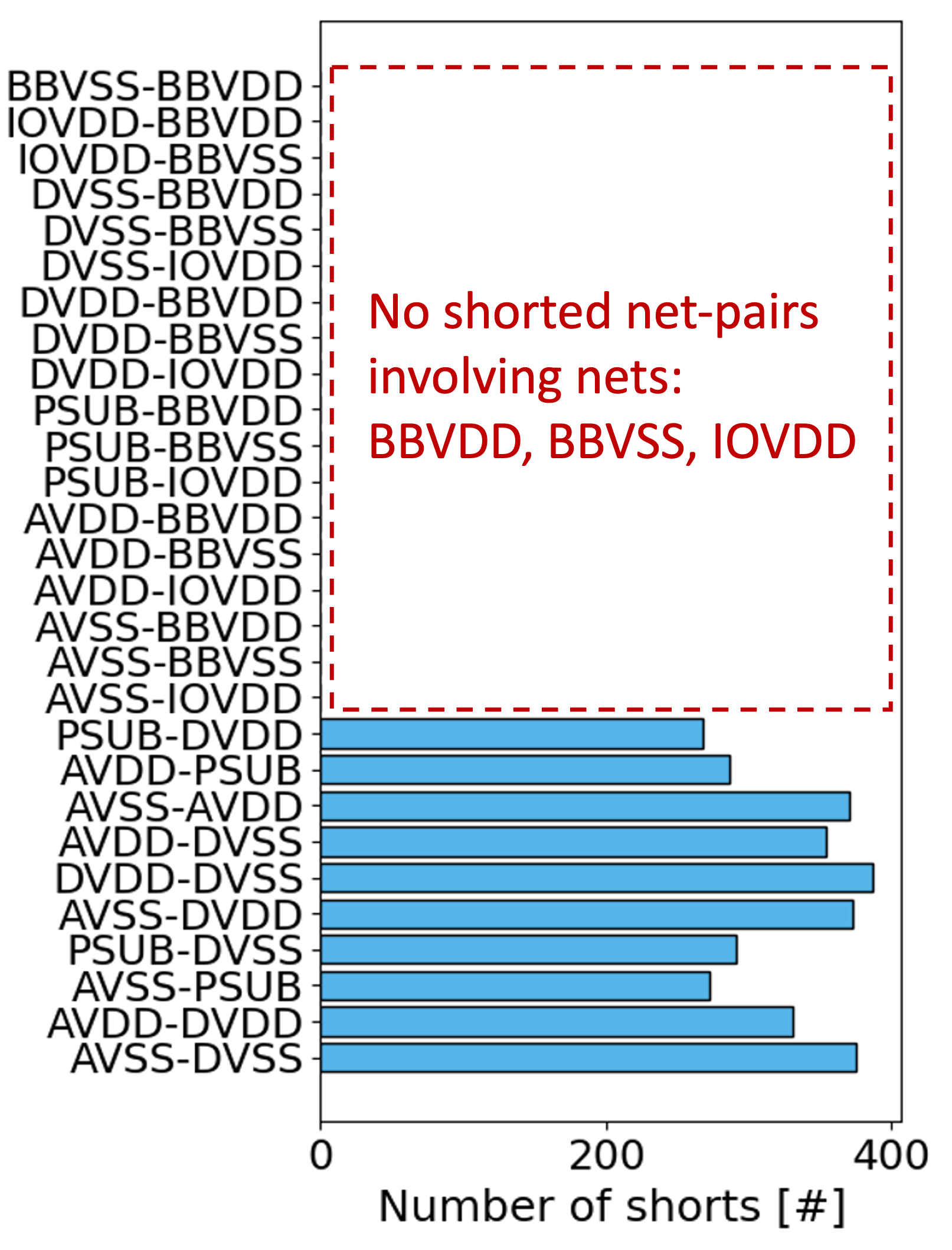}}
\caption{Shorts for all 28 net pair combinations. No shorts are observed measuring BBVDD, BBVSS, IOVDD nets.}
\label{fig:impedance_shorts_per_net}
\end{figure}

Shorts only occur in net combinations of nets AVDD, DVDD, AVSS, DVSS, PSUB as shown in Fig.~\ref{fig:impedance_shorts_per_net}. There is no favored net combination. No shorts are observed in net combinations involving BBVDD, BBVSS, IOVDD. Hence, only $\binom{5}{2}=10$ out of 28 power net combinations exhibit shorts.

The wafer-level distribution of the number of shorts per HU across the 20~wafers measured is given in Fig.~\ref{fig:shorts_per_HU_on_wafer} (see Supplementary Material and~\cite{Eberwein2025PhD}), with a higher number of shorts observed in the center of the wafers.

\subsection{Power ramping and hotspot location distribution}
\begin{figure}[hbt]
\centerline{\includegraphics[width=\linewidth]{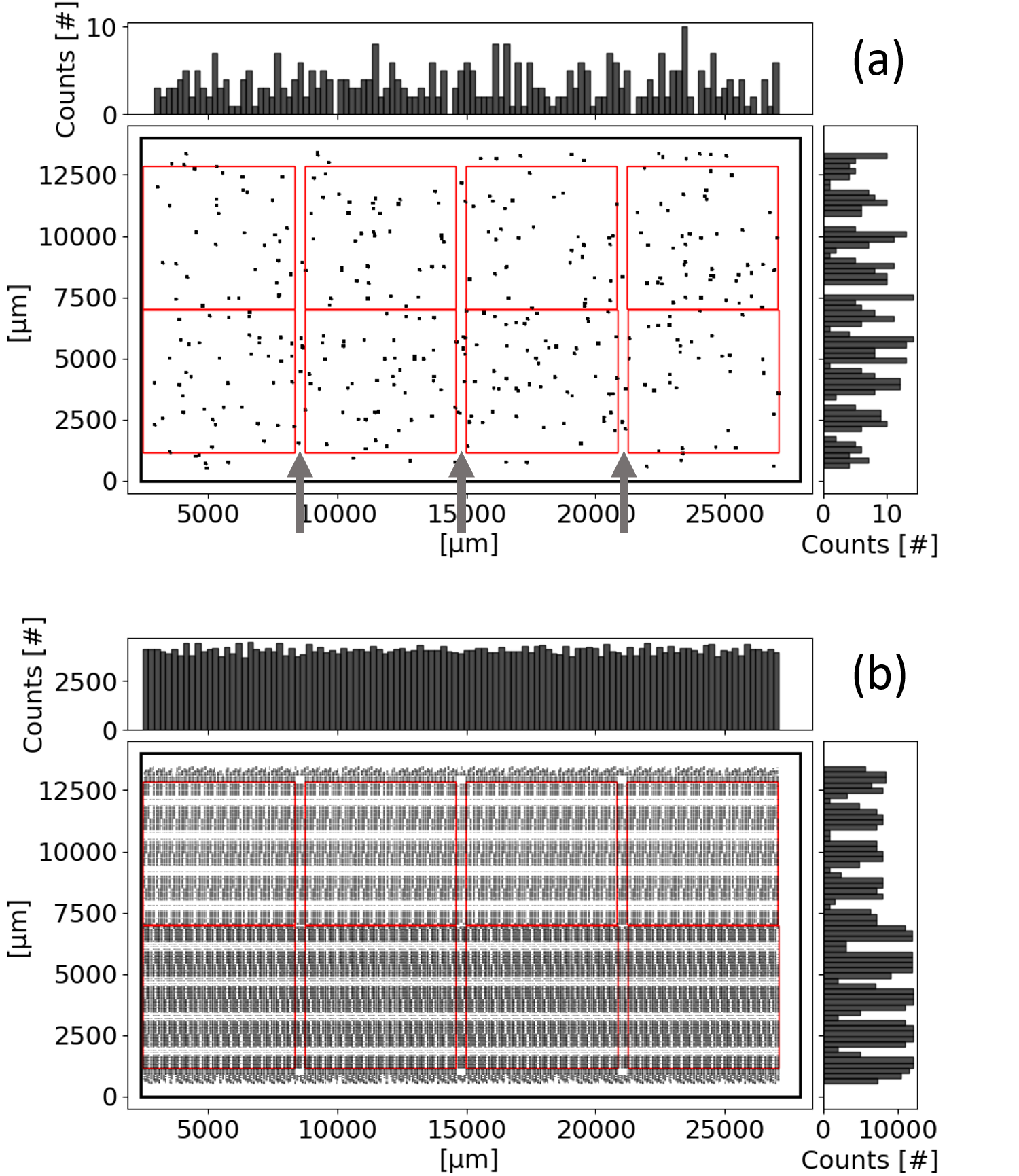}}
\caption{Measured hotspot locations (a), 
and areas with M7 and M8 present simultaneously (b). The arrows in (a) indicate the vertical gaps between pixel matrices (red squares) where only M7 and M8 metals are routed.
}
\label{fig:hotspot_locations}
\end{figure}

The distribution of a set of identified hotspots is mapped onto one RSU in chip coordinates as shown in Fig.~\ref{fig:hotspot_locations}a. Shorts are seemingly distributed without a distinct pattern. It is observed that shorts occur in regions between pixel matrices (indicated by arrows). Only the top two copper metals -- M7 and M8 -- are present in these regions.

From the chip design, regions with the presence of both M7 and M8 are extracted and shown in Fig.~\ref{fig:hotspot_locations}b. Here, these locations are shown for one RSU. 

The power ramp (PSUB~=~0~V) was measured for a total of 81 MOSS (1620~HUs) from 14 wafers with the following results:
\begin{itemize}
    \item 61.5\% show no transient high current,
    \item 34.2\% show a transient high current, corresponding to a burnt-through short,
    \item 4.3\% show a persistent high current or hotspot outside the operating limits.
\end{itemize}

HUs with burn-throughs are operated successfully in functional tests (such as reading and writing registers, and digital and analog pixel scans, performed on a dedicated test system~\cite{MOSS_first_characterization}), and no correlation between operating failures and burn-throughs was observed. Overall, 89\% of HUs with at least one short can be operated within specifications after the short is removed by supplying a sufficiently large current, resulting in a burn-through. The distribution of burn-through currents and voltages is shown in Fig.~\ref{fig:burn_through_currents} (see Supplementary Material). It is important to note that these currents are moderate (see also Table~\ref{tab:power_on_currents}) and do not pose any danger to the chip power supply network. In the remaining cases, the short remains, as an increase in current potentially burning through the short would only be possible by increasing the supply voltage above a safe level.

\subsection{Hypothesis formation}
Findings from data analysis are summarized below and allow us to form a hypothesis on the fault mechanism.
\begin{enumerate}
    \item Wafer-to-wafer fluctuations. Large statistical variations in the number of shorts, of up to a factor of 10, were observed between different wafers with hotspots in varying locations (see Fig.~\ref{fig:short_per_wafer}).
    \item Integration density independence. A comparable number of shorts are observed in the top and bottom halves of the MOSS chips (see Fig.~\ref{fig:short_per_wafer}),  designed with large and small line spacing (low and high integration density), respectively. The line spacing ratio of top/bottom ranges from 3/2 to 5/4.
    \item Shorts are also observed in the gaps with only M7 and M8 present in between pixel matrix areas (see arrows in Fig.~\ref{fig:hotspot_locations}a).
    During inspections of fault locations with a microscope, no optical evidence was found of shorts involving the presence of one metal only -- now referred to as Hypothesis B.
    \item The metal stack composition (both thickness and dielectric) for layers M7 and M8 is different from the remaining metal stack. 
    \item The fault locations extracted with the thermal camera match regions in the chip with specific features involving both M7 and M8 metals.
    \item No shorts are observed for 3 out of 8 power nets (BBVDD, BBVSS, IOVDD). M7--M8 metal features differ for these power nets compared to the rest of the power grid.

\end{enumerate}

The following Hypothesis A posits that shorts correlate with areas that have specific layout features involving both M7 and M8. The alternate Hypothesis B posits that shorts correlate with specific layout areas involving one metal only (M7 or M8).

\section{Hypothesis validation and implications}
Following the procedure described above, the single hotspot locations are correlated with the corresponding single low impedance net pair and the chip design coordinates. A sample of 156 such instances was analyzed. 
We check if the affected net pair M7--M8 metals lie within the resolution window and are compatible with Hypothesis A. Compatibility with Hypothesis B is additionally tested for. 
The results are summarized in Table~\ref{tab:crossing}. Excellent agreement with Hypothesis A over Hypothesis B is observed. For the 100~$\upmu$m$\,\times\,$100~$\upmu$m resolution window, 147/156 (94\%) test cases are in agreement with Hypothesis A. Manual case-by-case analysis of the remaining non-compatible occurrences found that 155/156 (99\%) of test cases agree with Hypothesis A, when increasing the resolution window by 50~$\upmu$m (equivalent to a 1--pixel shift).

\begin{table}[h]
    \centering
    \caption{\textsc{Hypothesis Compatibility}}
    \begin{tabular}{| >{\centering\arraybackslash}m{1.8cm} | >{\centering\arraybackslash}m{1.8cm} | >{\centering\arraybackslash}m{1.8cm} | >{\centering\arraybackslash}m{1.1cm} |}
         \hline
         Resolution window [$\upmu$m$^2$] & Hypothesis A compatible & Hypothesis B compatible & Total counts\\
         \hline 
         \hline
         50 $\times$ 50 & 89\% & 16\% & 156\\
         100 $\times$ 100 & 94\% & 31\% & 156\\
         \hline
    \end{tabular}
    \label{tab:crossing}
\end{table}

From the chip design, we also extract the total layout areas compatible with Hypothesis A, and observe an even area split between the top (49.8\%) and bottom (50.2\%) halves of the chip. This matches the integration density independence, observed in Fig.~\ref{fig:short_per_wafer} as an even split of shorts in the top and bottom halves of the chips.

\subsection{Cross-section imaging}
Using Focused Ion Beam-Scanning Electron Microscopy (FIB-SEM), it is possible to create a cross-section image at the expected short location. An initial cut is made with a Ga-Ion beam 50~to~100~$\upmu$m from the expected fault location, and gradually advanced. The cross-section image is monitored with the scanning electron microscope. Two separate samples were analyzed: before and after burning through a short, respectively.

In sample SA01, the power ramp was stopped once the hotspot and thus the fault location was identified, but before the short was burnt through. This was confirmed by an impedance measurement, indicating the short was still present. A short structure consistent with Hypothesis A was found. Energy Dispersive X-Ray Spectroscopy (EDS) was used to confirm that the connecting structure is copper. 

An additional cross-section analysis was performed on sample SA02, where the short was burnt through, and the impedance before and after powering the chip changed from low to high. The resulting cross-section image exhibits a small cavity formation around the fault, effectively breaking the short. The structure itself is consistent with Hypothesis A. 
During the measurements, it has been observed that a burnt-through short sometimes reconnects, manifesting itself in a current rise and the reappearance of the hotspot (see Fig.~\ref{fig:double_burn_through} in the Supplementary Material). A plausible explanation is the local expansion and contraction of the metal stack caused by the highly localized temperature change introduced by the short fault.

\subsection{Mitigation and future testing}
The analysis clearly established that the shorts were associated with layout features involving both M7 and M8 metals -- new metal layers introduced for the MOSS chip by the foundry in this collaborative effort. Together with the cross-sectional images, this feedback enabled the foundry to implement a mitigation strategy, including the recommendation of revised design rules, to eradicate the observed failure mode in future sensors.

Given that burn-throughs occur at currents comparable to or below chip operating currents, the observed failures would have been missed if the chip had been powered on without measuring impedances, carefully ramping up the power nets, and using a thermal camera. This has important implications for future chip characterization campaigns. The initial impedance measurement and power-ramping steps will be kept, allowing for observing short faults otherwise potentially masked, and understanding and disentangling the contributions to yield loss originating in the chip metal stack.

\section{Conclusion}
To assess the yield for stitched sensors for ALICE ITS3, MOSS sensors were characterized. The metal stack integrity was tested using a novel approach. A single lot of 24 wafers was manufactured in an experimental engineering run, using a custom metal stack composition introduced in a collaborative effort with the foundry. Short faults were observed on all 20 tested wafers with varying frequencies. Dedicated impedance and powering setups were developed, including the use of a thermal camera for fault localization. In-depth data analysis revealed the root cause to be shorts involving the top two copper metal layers, collaboratively introduced in this chip. This was further confirmed using FIB-SEM cross-section analysis. 89\% of shorts were observed to be burnt through by moderate, sufficiently large currents, after which these chips can be operated successfully. Without impedance measurements and slow ramp-up of power nets during initial power-up (including the use of a thermal camera), these shorts would have been missed. Some of the burnt-through shorts were also observed to reconnect or reappear. The feedback from this analysis and the cross-section imaging allowed the foundry to implement a mitigation strategy, as well as provide adapted design rules, to avoid this failure mode in future fabrication runs. The correlation of impedance, power-ramp-up, and thermal camera measurements, along with chip layout information, proves to be a powerful approach for root cause defect analysis and is generally applicable to CMOS devices with advanced interconnect technology.

\bibliographystyle{IEEEtran}  
\bibliography{bibliography_2}

\vspace{8.5cm}
\section*{Supplementary material}
Supplementary figures referenced in the text are given here.

\begin{figure}[b]
\centerline{\includegraphics[width=\linewidth,trim={0 0 0 0.35cm},clip,keepaspectratio]{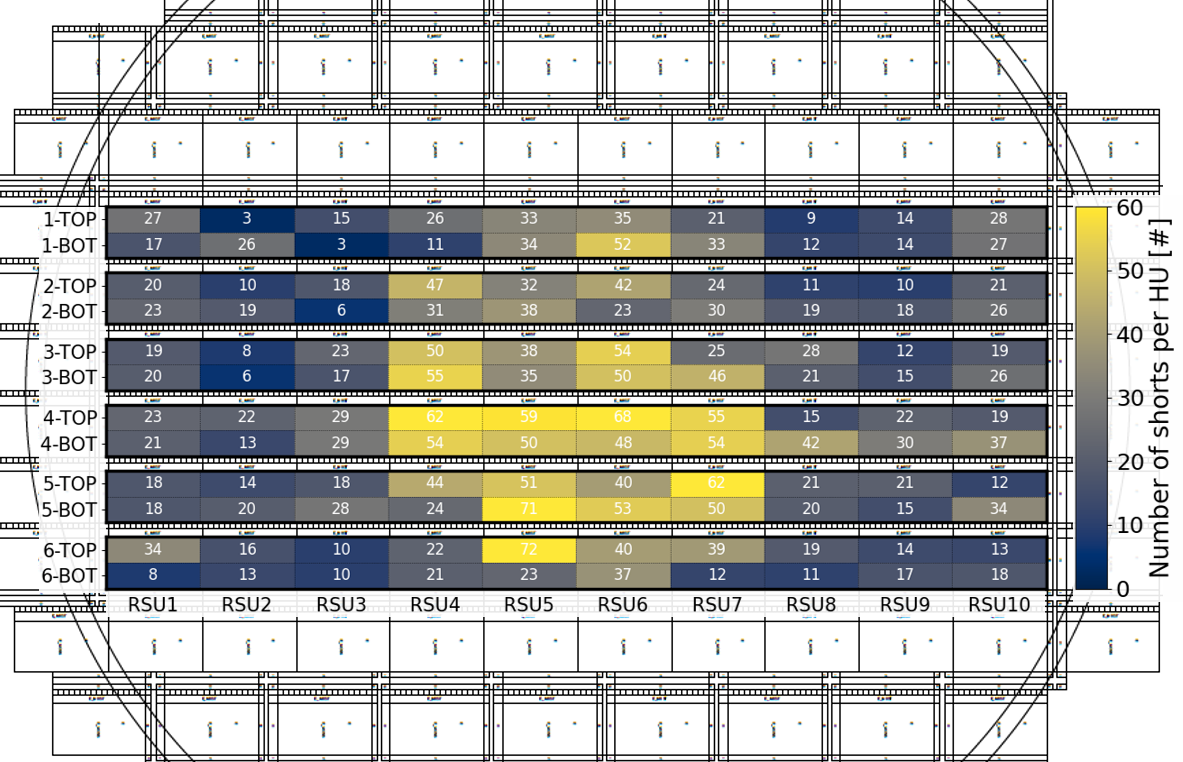}}
\caption{Distribution of number of shorts per HU on wafer level. Each double row represents the position of one MOSS sensor on the wafer, with 20 HUs per chip. A distinct central gradient is visible. This effect is attributed to the type of failure observed and is compatible with high-temperature and rotationally symmetric processing steps during fabrication.}
\label{fig:shorts_per_HU_on_wafer}
\end{figure}

\begin{figure}[h]
\centerline{\includegraphics[width=\linewidth]{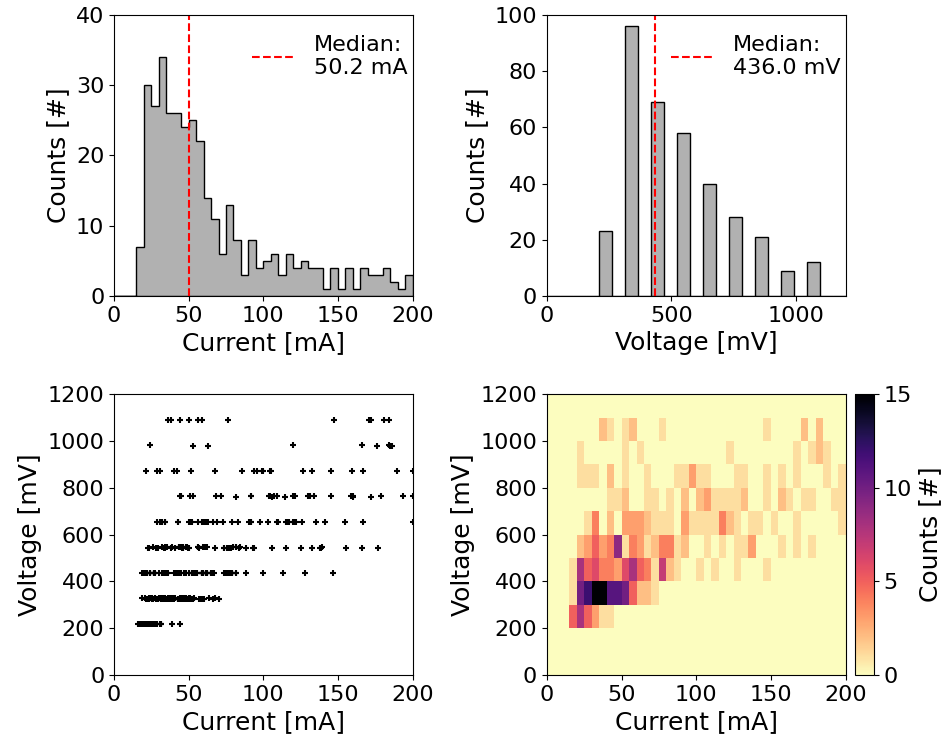}}
\caption{Distribution of burn-through currents and voltages. The discrete voltage steps correspond to the chosen steps during chip power ramp-up. }
\label{fig:burn_through_currents}
\end{figure}

\begin{figure}[!t]
\centerline{\includegraphics[width=\linewidth]{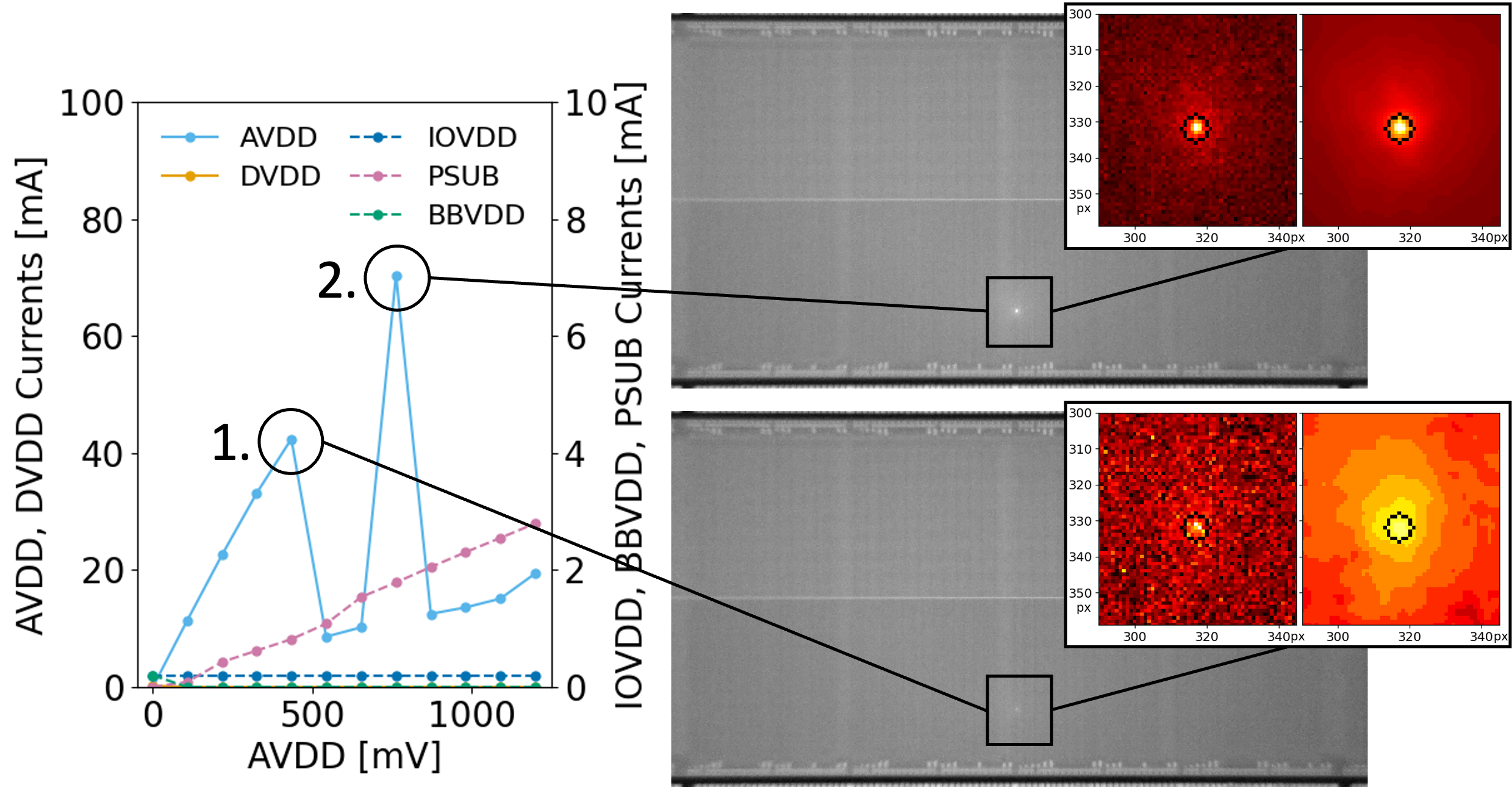}}
\caption{Example of a double burn-through: The short disappears temporarily (first peak), before burning through permanently (second peak). The same hotspot location is identified for both peaks, and magnified figures are overlaid showing the non-denoised and denoised analysis step for each peak. The local temperature change during the first burn-through appears to cause deformation in the metal stack, resulting in the short reconnecting. Only after the second burn-through does the impedance change from low to high and stay stable. The increase in PSUB current is within specification.}
\label{fig:double_burn_through}
\end{figure}

\end{document}

%% file: ieee_author_block.tex
\author{%
Gregor Hieronymus Eberwein,
Gianluca Aglieri Rinella,
Daniela Bortoletto,
Szymon Bugiel,
Francesca Carnesecchi,
Antonello Di Mauro,
Pedro Vicente Leitao,
Hartmut Hillemanns,
Marc Alain Imhoff, 
Antoine Junique,
Alex Kluge,
Magnus Mager,
Paolo Martinengo,
Iaroslav Panasenko,
Ivan Ravasenga,
Felix Reidt,
Valerio Sarritzu,
Walter Snoeys,
Miljenko Šuljić

\thanks{Manuscript received xx.xx.2025, .. }
\thanks{G. H. Eberwein was with the University of Oxford, OX1 2JD Oxford, UK. He is now with the Massachusetts Institute of Technology, Cambridge, MA 02139, USA (e-mail: eberwein@mit.edu).}
\thanks{G. Aglieri Rinella, S. Bugiel, F. Carnesecchi, A. Di Mauro, P. V. Leitao, H. Hillemanns, A. Junique, A. Kluge, M. Mager, P. Martinengo, I. Ravasenga, F. Reidt, W. Snoeys, and M. Šuljić are with the European Organization for Nuclear Research (CERN), 1211 Geneva, Switzerland.}
\thanks{D. Bortoletto is with the University of Oxford, OX1 2JD Oxford, UK.}
\thanks{M. A. Imhoff is with the Centre National de la Recherche Scientifique, 67200 Strasbourg, France.}
\thanks{I. Panasenko is with Lund University, 221 00 Lund, Sweden.}
\thanks{V. Sarritzu is with Universit\`a degli studi di Cagliari, 09124 Cagliari, Italy and the Istituto Nazionale di Fisica Nucleare (INFN), Sezione di Cagliari, Italy and the European Organization for Nuclear Research (CERN), 1211 Geneva, Switzerland.}

}